# Quantum Tunneling in the Surface Diffusion of Single Hydrogen Atoms on Cu(001)

Xiaofan Yu[1,2] (于小凡), Yangwu Tong[1,2] (童洋武) and Yong Yang[1,2*](杨勇)

1. *Key Lab of Photovoltaic and Energy Conservation Materials, Institute of Solid State Physics, HFIPS, Chinese Academy of Sciences, Hefei 230031, China.*
2. *Science Island Branch of Graduate School, University of Science and Technology of China, Hefei 230026, China.*

**ABSTRACT:**

The adsorption and diffusion of hydrogen atoms on Cu(001) are studied using first-principles calculations. By taking into account the contribution of zero-point energy (ZPE), the originally identical barriers are shown to be different for H and D, which are respectively calculated to be ~ 158 meV and ~ 139 meV in height. Using the transfer matrix method (TMM), we are able to calculate the accurate probability of transmission across the barriers. The crucial role of quantum tunneling is clearly demonstrated at low-temperature region. By introducing a temperature-dependent attempting frequency prefactor, the rate constants and diffusion coefficients are calculated. The results are in agreement with the experimental measurements at temperatures from ~ 50 K to 80 K.

**Keywords:** H/Cu(001), first-principles calculations, quantum tunneling, diffusion coefficients

*Corresponding author: yyanglab@issp.ac.cn

## 1. Introduction

Diffusion of adsorbates is a fundamental process for reactions on metal surfaces. [1,2] The diffusion of hydrogen atoms on metal surfaces is an important step in many catalytic reactions, such as the formation of ammonia in the Haber-Bosch process, [3] the synthesis of hydrocarbons, [4] and the generation and oxidation of $H_2$ in electrochemistry. [5,6] Numerous experimental [7-12] and theoretical [13-19] studies have been carried out to understand the process of hydrogen diffusion on solid surfaces. Experimental techniques such as helium atom scattering (HAS), [9] scanning tunneling microscopy (STM), [10,11] inelastic electron tunneling spectroscopy (STM-IETS), [10,11] high-resolution electron energy loss spectroscopy (HREELS), [20] and linear optical diffraction (LOD) techniques [7] have been used to study the quantum effects of hydrogen atoms during diffusion. Quantum mechanics is applied to deal with the quantum motions of atomic nuclei by *ab initio* path integral molecular dynamics [21] and transition state theory (generalized instanton theory) based on Feynman path integrals. [22,23] Density functional theory (DFT) is typically used to calculate the three-dimensional adiabatic potential energy surface (PES) for hydrogen diffusion on metal surfaces.[13,14] It is generally accepted that the hydrogen diffusion path on the Cu(100) surface follows a quadruple hollow site across bridge to an adjacent quadruple hollow site. [13] Lauhon and Ho [10,11] used scanning tunneling microscopy (STM) to directly measure the hopping rates of individual hydrogen and deuterium atoms on the Cu(001) surface at low temperatures and low coverage. Their study shows that the diffusive motion of hydrogen atoms undergoes a transition from thermal activation-dominated to quantum tunneling-dominated at a temperature of about 60 K. Basically, the potential barrier along the diffusion path determines the rate constant and diffusion coefficient for surface diffusion. To date, theoretical studies are still in disagreement on the height of the diffusion potential barrier. Lauderdale and Truhlar treat surface diffusion of H on Cu(001) by a model involving 21 degrees of freedom and obtained a barrier of 507 meV. [24] Kua *et al.* obtained a barrier height of 175 meV by DFT calculations. [25] Sundell and Wahnström obtained a diffusion barrier height of 126 meV by considering adsorbed hydrogen atoms with a coverage of 1/9 and a barrier height of 180 meV after zero-point energy correction.[14,15] Lai *et al.* found that the minimum of the potential is located at the fourfold hollow site with a diffusion barrier of 88 meV at the bridge site based on DFT calculations and a three-dimensional spline interpolation.[13] The difference in the potential barrier height could be attributed to the variation in the methods employed to model the H-atom-surface interaction. Kua *et al.*[25] did not take into account the structural relaxation of the surface Cu atoms during adsorption and diffusion, while Sundell and Wahnström[14,15] considered the effect of structural relaxation.

At sufficiently low temperature, the classical diffusion of small-mass adsorbates such as hydrogen gives way to the process dominated by quantum tunneling. Therefore, quantum tunneling effects must be explicitly considered during theoretical calculations. Sundell and Wahnström[15] calculated the rate constants by considering the effect of lattice relaxation (hydrogen distorting the positions of surrounding metal atoms) and the nonadiabatic response of the conduction electrons to the hydrogen motion. However, their results differed from experiment[10,11] by an order of magnitude in the low temperature region.

In this paper, we revisit this topic to study the quantum effects on the diffusion of H/D atoms on Cu(001) surface at low temperatures. First, we determined the diffusion potential barrier based on first-principles calculations. Then, we used the transfer matrix (TM) method to investigate the quantum tunneling effect. We calculated the phonon spectrum of the system containing hydrogen and surface copper atoms and obtained the vibrational frequencies at typical sites are comparable with previous studies. The diffusion potential corrected by the zero-point energy is obtained in this study. To calculate the diffusion rate constants and diffusion coefficients of H/D atoms on Cu(001), we introduced a spectrum-dependent attempting frequency prefactor. At temperatures between 50 K and 80 K, the theoretical results are in good agreement with those measured experimentally.[10,11]

## 2. Methods

### 2.1 Details of First-principles Calculations

The DFT calculations are carried out using the Vienna *ab initio* simulation package (VASP). [26,27] The Perdew-Burke-Ernzerhof (PBE) type functional [28] and dispersion-corrected density functional theory (DFT-D2) [29] are used to describe the exchange-correlation terms of electrons, in combination with the PAW potentials [30,31] to describe the electron-ion interactions. The energy cutoff for plane wave basis sets is 600 eV. The initial atomic configurations are constructed with the aid of VESTA, [32] in which the Cu(001) surface is modeled by a six-layer *p*(3×3) supercell, repeating periodically along the *xy* plane with a vacuum layer of about 15 Å along the *z* direction. In all the calculations, the Cu atoms in the bottom three layers are fixed and the atoms in the upper layers are relaxed. We employ a dipole correction for the total energy to eliminate the artificial dipole-dipole interaction caused by the upper and lower asymmetric slab surfaces. A 4×4×1 Monkhorst-Pack k-mesh [33] is generated for sampling the Brillouin zone (BZ) in performing structural relaxation and total energy calculations. The adsorption energy for atomic hydrogen is given by

$$E_{ads} = E_{H/Cu(001)} - E_{Cu(001)} - E_H \tag{1}$$

where $E_{H/Cu(001)}$ and $E_{Cu(001)}$ are the energies of the Cu(001) system with and without the H adsorbate. $E_H$ is the ground state energy of the free H atom (in the spin-polarized state).

We used the climbing image nudged elastic band (CI-NEB) method [34] to calculate the minimum energy path. The vibrational properties of the relaxed structure are analyzed by density functional perturbation theory (DFPT). [35] The vibrational spectrum and the corresponding phonon density of states for each adsorption site are obtained by DFPT calculations carried out using VASP jointly with the Phonopy package. [36] The zero-point energy (ZPE) is then obtained by summing up the contribution of all the vibrational modes. The energy pathway of diffusion is then corrected by adding the zero-point energy to the corresponding atomic configurations.

**2.2 The Transfer Matrix Method**

The transfer matrix method (TMM) is a numerically accurate and efficient method for dealing with the transmission of quantum particles across a given potential field. [37, 38] This is evident by comparing to results of TMM [37,38] with that obtained by the semi-classical Wentzel-Kramers-Brillouin (WKB) approximation.[39] When the energy of a quantum particle is less than the height of the barrier, the particle has the probability of passing through the barrier due to the quantum tunneling effect. When the energy of the quantum particle is larger than the barrier height, due to the existence of quantum interference, there may be a certain probability of reflection which makes the probability of passing through the barrier less than 1. The WKB approximation only considers the quantum tunneling effect and does not account for quantum interference. To perform TM calculations, we numerically slice the one-dimensional potential barriers along the direction of hydrogen atom diffusion to obtain multiple small rectangular-like potential barriers. When the particle passes through the potential barrier, it can be viewed as passing through multiple small rectangular barriers in succession, and each transmission can be represented by a coefficient matrix describing the transmission and reflection amplitudes of the wave function. Multiplying the coefficient matrices sequentially yields a transfer matrix that represents the transition relationship between the initial and final states. The expressions are as follows [37,38]:

$$\begin{pmatrix} A_R \\ B_R \end{pmatrix} = M \begin{pmatrix} A_L \\ B_L \end{pmatrix} = \begin{pmatrix} m_{11} & m_{12} \\ m_{21} & m_{22} \end{pmatrix} \begin{pmatrix} A_L \\ B_L \end{pmatrix} \quad (2)$$

where $A_L$ and $B_L$ are the incoming amplitudes, and $A_R$ and $B_R$ are the outgoing amplitudes. In a system which preserves the time-reversal symmetry, the determinant $|M| = 1$，and the transmission coefficient is calculated by $T_r(E) = \frac{1}{|m_{22}|^2}$.

## 3. Results & Discussion

Firstly, we calculated the adsorption energies ($E_{ads}$) of single H atom on top site (TP), bridge site (BR), and hollow site (HL) in Fig.1 using DFT-PBE and DFT-D2, respectively, and deduced their zero-point energies (ZPE) by DFPT calculations. The results are placed in Table I. We focus on two typical diffusion paths, namely path 1 (hollow-bridge-hollow) and path 2 (hollow-top-hollow) as illustrated in Fig. 1. The calculated potential barriers using CI-NEB are given in Fig. 2(a) and 2(b), respectively. It is evident from the barrier height (~ 0.12 eV vs ~ 0.6 eV) that the hydrogen diffusion process will primarily follow path 1. In the following studies, our attention will focus on the diffusion along path 1.

Zero-point vibration and the related zero-point energy (ZPE) are among the key features that distinguish a quantum particle from a classical one. A part of the zero-point energy data in Table I are calculated by DFPT containing only the Γ point, i.e., phonons of the long wavelength limit for which the wave vectors $q = 0$. However, this approximation is only strictly applicable to the study of gaseous states or atomic and molecular vibrational modes where the vibrations are dominated by a few number of isolated frequencies. For surface adsorption, where the vibration is a continuous spectral distribution due to the stronger coupling motion of the adsorbed atoms and the substrate, a more accurate zero-point energy of hydrogen on the Cu(001) surface needs to be calculated by a weighted sum of all the vibrational modes as follows:

$$ZPE = 0.5 \times h \int_0^{v_{max}} v g(v) dv \tag{3}$$

Where $v_{max}$ is the phonon maximum frequency, which roughly corresponds to the Debye frequency, $g(v)$ is the vibrational density of states at frequency $v$, and $h$ is the Planck constant.

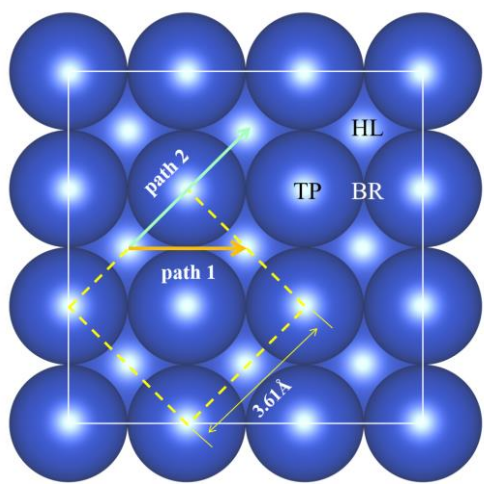

**Fig. 1** Schematic diagram of hydrogen diffusion paths on Cu(001). Path 1 is the diffusion path following the hollow site across a bridge site to an adjacent hollow site; path 2 is the diffusion path following the hollow site across a top site to an adjacent

hollow site. The abbreviation HL denotes the hollow site, TP for the top site, and BR is the bridge site.

**Table I.** Calculated adsorption energies ($E_{ads}$) and zero-point energies (ZPE) of H atoms on Cu(001), and the H-Cu bond lengths ($d_{CuH}$), H-Cu(001) distance ($Z_{CuH}$). For each quantity, the results obtained by PBE and DFT-D2 calculations are in the upper and lower lines, respectively. In PBE calculations, the ZPE obtained by Γ-only and $q$-dependence (data in parentheses) DFPT calculations are listed for comparison.

|  | Top | Bridge | Hollow |
|---|---|---|---|
| Eads (eV) | 1.673 | 2.146 | 2.276 |
|  | 1.743 | 2.231 | 2.348 |
| ZPE (eV) | 0.108 (0.107) | 0.149 (0.146) | 0.111 (0.113) |
|  | 0.111 | 0.151 | 0.109 |
| $d_{CuH}$ (Å) | 1.52 | 1.65 | 1.87 |
|  | 1.51 | 1.64 | 1.88 |
| $Z_{CuH}$ (Å) | 1.55 | 1.13 | 0.55 |
|  | 1.51 | 1.09 | 0.56 |

The calculated ZPE is ~ 0.11 eV at the hollow site, which is in good agreement with the results of Sundell and Wahnström [14,15]. The ZPE results including the contribution of the phonons with wave vectors $q > 0$ are given inside parentheses of Table I. The small difference of computed ZPE implies that DFPT calculations involving only the Γ point are usually sufficient to describe the vibrations of H on Cu(001). Meanwhile, it is also seen that the accuracy of the DFT-PBE description of the H-Cu(001) interaction is sufficient by comparing the results of DFT-PBE with that of DFT-D2 which includes the effects of van der Waals forces.

Due to the strong coupling between the adsorbed atoms and the substrate Cu atoms, the ZPEs of all atoms in the model system were calculated in order to obtain more accurate correction to the diffusion barrier heights for the H/D atoms on the Cu(001) surface. Let $x$ denote the coordinates along the diffusion path, then the ZPE-corrected potential $V(x) = V_0(x) + E_0(x)$, with $V_0(x)$ the original (uncorrected) potential and $E_0(x)$ the ZPE at position $x$. The diffusion potential barrier along path 1 with and without ZPE corrections are shown in Fig. 2(a) for single H and D atoms. As a result of different vibrational frequencies at different adsorption sites, in particular, the ZPE difference at the hollow and bridge sites (~ 0.113 vs 0.146 eV for H, Table I), the ZPE-corrected barrier height is significantly higher. Furthermore, we found that H and D atoms have different potential barrier heights due to their different masses and consequently vibrational frequencies and the corresponding ZPE. Within the harmonic approximation, the vibrational frequency is proportional to the inverse of square root of particle mass, i.e., $v \propto \frac{1}{\sqrt{m}}$. It turns out that the height of the diffusion barrier of single H atoms on the surface of Cu(001) is ~ 0.158 eV, and the height of the diffusion barrier of single D atoms on the surface of Cu(001) is ~ 0.139 eV. Due to their larger particle mass, the ZPE-correction ($\Delta E_0$) on the diffusion path of single D atoms is smaller than that of H by a factor of $\sqrt{2}$: $\Delta E_{0,D}(x) = \frac{1}{\sqrt{2}}\Delta E_{0,H}(x)$. This explains the difference between the ZPE-corrected barriers (Fig. 2(a)) for H and D diffusion on Cu(001).

The barrier height after ZPE-correction for H/D atomic surface diffusion in Sundell and Wahnström's work [14,15] is 180/170 meV, whereas the barrier heights in their work are essentially the same as that obtained in our work prior to taking into account the ZPE correction, which is ~ 0.13 eV. The difference in barrier height is primarily attributed to the values of ZPE corrections. The reason for this discrepancy may be the difference of the computational accuracy. Using the same VASP code for DFT calculations, the energy cutoff for plane wave basis sets in our calculation is 600 eV while that of Sundell and Wahnström is about 272 eV (20 Ry). Additionally, the impact of the spectral distribution of substrate atomic vibrations was not considered in Sundell and Wahnström's work. The coupling of vibrational motions between the adsorbed H/D atoms and the substrate Cu atoms is nontrivial, it is important to consider contributions from all the atoms in the system to obtain more accurate ZPE corrections.

Figures 2(c)-(d) shows the phonon spectra of H adsorption at the hollow and bridge sites and the corresponding vibrational density of states. From the density of states, it can be seen that the vibrations at energies of ~ 35 meV and below are mainly from the lattice vibrations of Cu atoms, while the high frequency part (energies of ~70 meV and above) mainly corresponds to the vibrations of the adsorbed H atoms. For the hollow site, the vibrational mode (~ 80 meV) in the perpendicular direction is slightly larger while the two degenerate vibrational modes (~ 70 meV) in the direction parallel to Cu(001) are in good agreement with the experimentally value of ~ 70 meV.[12] It is also seen that the vibrational frequencies of the H atom at the bridge site along the $y$ and $z$ directions are nearly degenerate with the phonon energy of ~ 140 meV. In addition, the presence of a notable imaginary frequency peak in the phonon spectrum of the bridge site indicates that the vibrational mode along the $x$-direction (the direction of diffusion along path 1) is dynamically unstable.

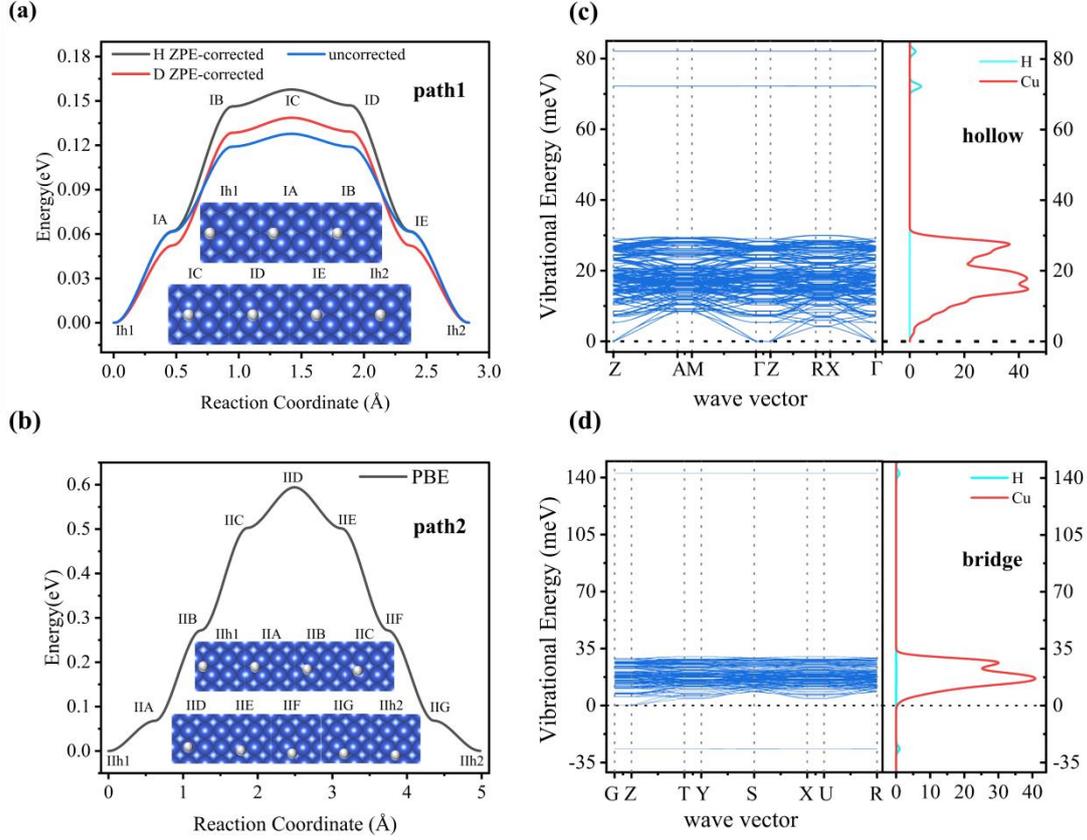

**Fig. 2** Left panels: Energy barriers for the diffusion of H/D along the two paths. For path 1, the energy barriers with and without ZPE corrections are displayed. Right panels: Phonon spectra of H/Cu(001) at the hollow and bridge site.

To study the kinetic properties of diffusion on the Cu(001), we calculate the transmission coefficient $T_r(E)$ for H/D atoms crossing the potential barrier in path 1 (Fig. 2(a)) at a given energy $E$ by the transfer matrix (TM) method and the WKB method, respectively. Considering the fact of local thermal fluctuations, under the condition of thermal equilibrium, the degrees of freedom of translational motions of the adsorbed H/D are excited or partially excited. Then its kinetic energy $E$ at temperature $T$ satisfies the distribution[37] $p(E,T) = 2\pi(\frac{1}{\pi k_B T})^{3/2}\sqrt{E}e^{-E/k_B T}$, where $k_B$ is the Boltzmann constant. The total transmission probability is given by:

$$P_{tot}(T) = \int_0^{E_{max}} p(E,T)\, T_r(E)\, dE \tag{4}$$

where the upper limit of integration is the energy $E_{max} = h\nu_{max-bridge}$,

corresponding to the maximum phonon frequency of H/D adsorption at the bridge site.

The role of quantum effect can be demonstrated by assuming that H/D atoms are classical particles, for which barrier-crossing is possible only when their kinetic energies are larger than the barrier height. Therefore, the probability corresponding to this process can be calculated as follows:

$$P_C(T) = \int_{E_b}^{E_{max}} p(E,T)dE \tag{5}$$

As above, $p(E,T) = 2\pi(\frac{1}{\pi k_B T})^{\frac{3}{2}}\sqrt{E}e^{-E/(k_B T)}$ is the kinetic energy distribution function at a given temperature $T$. $E_b$ is the height of the original diffusion barrier without ZPE-correction. Figure 3(a)-(b) gives the transmission coefficients $T_r(E)$ of H/D atoms calculated by TM and WKB. It is seen from the curves that the transmission coefficients of H/D oscillate between 0.15 eV and 0.3 eV of incident energy. This is due to the quantum interference of the incident and reflected matter waves of H/D atoms, while such phenomenon is absent in the semi-classical WKB method. Figure 3(c) and 3(d) show the total transmission probability of H/D as a function of temperature, calculated using the transmission coefficients obtained from full quantum method (TMM), semi-classical method (WKB), and classical method. The insets of Figs. 3(c)-3(d) compare the total transmission probabilities calculated by TMM, WKB and classical method at $T \leq 60$ K. At low-temperature region TMM and WKB predict nearly the same $P_{tot}(T)$, while the classical method gives much smaller values in the order of magnitude of several tens to hundred. For details, the total transmission probabilities obtained by different methods at $T =$ 10, 20, 30, 40 and 60 K are presented in Table II. The difference is clearly seen. At $T \leq 30$ K, the transmission probabilities taking into account the effects of quantum tunneling are much larger than that treating H/D as classical particles. The temperature is different for H and D when the quantum probability ($P_Q$) is equal to the classical probability ($P_C$): It is ~ 40 K for H while ~ 30 K for D. This is due to the quantum behavior caused by the difference in the mass of H and D. The smaller the mass, the more obvious the quantum behavior is. Due to the presence of quantum interference, a certain probability of reflection exists even if the energy of the quantum particle is larger than the height of the potential barrier, which is more remarkable for lighter atoms. The results obtained by TM calculations treat this phenomenon accurately, since the TM method fully takes into account the quantum effects in the process of crossing the potential barrier. Therefore, the results of TM calculations are more accurate than those of the WKB method.

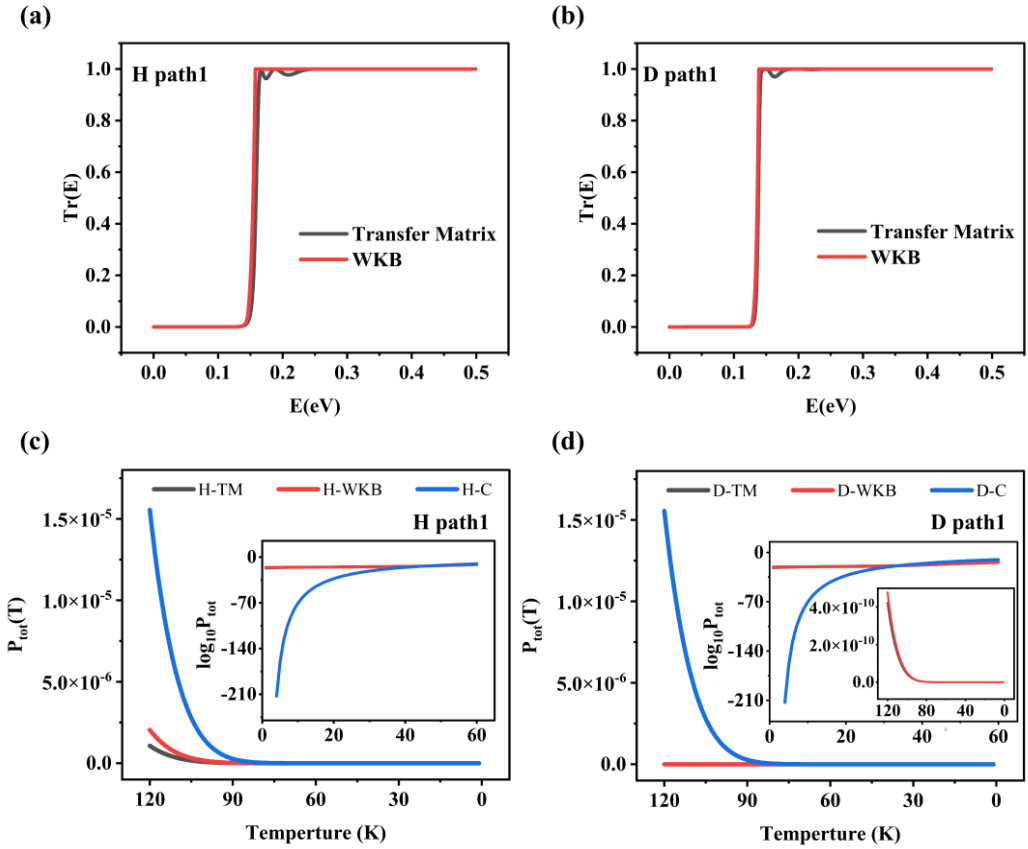

**Fig. 3** Calculated transmission coefficients of H/D by TM method and WKB along Path 1 (Panels (a)-(b)) and the corresponding total transmission probabilities as a function of temperature (Panels (c)-(d)), with the data for $T \leq 60$ K highlighted in the insets. The total transmission probabilities of a classical particle are shown for comparison. The data labeled by H-C (D-C) is the total transmission probability obtained by treating the H (D) atom as a classical particle.

On the other hand, the classical probability of transmission increases fast with increasing temperature. As shown in Fig. 3, from $T \sim 30$ to 80 K, the classical probability ($P_C$) has nearly the same values as the quantum probability ($P_Q$). From $T \sim$ 80 K and above, the classical probability shows significantly larger magnitude than the quantum probability. This can be understood as follows. Assuming H/D atoms as classical particles, the ZPE correction to the potential barrier and the mass difference of the atoms are neglected. The large reduction in the height of the potential barrier leads to a significant overestimation of the classical transmission probability at relatively high temperature region.

**Table II.** Total transmission probability ($P_{tot}$) calculated by the TM method (TMM), WKB and classical method at different temperatures. The results for H and D are listed in the upper and lower lines, respectively.

| Temperature (K) | TMM | WKB | Classical |
|---|---|---|---|
| 10 | $2.97 \times 10^{-16}$ | $3.50 \times 10^{-16}$ | $4.14 \times 10^{-71}$ |
|    | $4.12 \times 10^{-21}$ | $4.32 \times 10^{-21}$ | $4.14 \times 10^{-71}$ |
| 20 | $9.46 \times 10^{-16}$ | $6.23 \times 10^{-16}$ | $1.30 \times 10^{-33}$ |
|    | $1.61 \times 10^{-20}$ | $9.88 \times 10^{-21}$ | $1.30 \times 10^{-33}$ |
| 30 | $2.64 \times 10^{-15}$ | $1.40 \times 10^{-15}$ | $5.10 \times 10^{-22}$ |
|    | $6.77 \times 10^{-20}$ | $4.28 \times 10^{-20}$ | $5.10 \times 10^{-22}$ |
| 40 | $9.27 \times 10^{-15}$ | $6.14 \times 10^{-15}$ | $2.18 \times 10^{-16}$ |
|    | $1.24 \times 10^{-18}$ | $1.55 \times 10^{-18}$ | $2.18 \times 10^{-16}$ |
| 60 | $3.38 \times 10^{-12}$ | $4.70 \times 10^{-12}$ | $7.03 \times 10^{-11}$ |
|    | $6.23 \times 10^{-15}$ | $6.84 \times 10^{-15}$ | $7.03 \times 10^{-11}$ |

For the reaction rate constant (i.e., hopping rate in case of surface diffusion), an important parameter is the attempting frequency or frequency prefactor. The frequency prefactor $v_{at}$ is directly related to the surface motion of the adsorbed atoms. The actually activated vibrational modes of the adsorbed H/D atoms at different temperature are different, so the frequency prefactor $v_{at}$ should be temperature dependent. As a try, we take the weighted average of all the vibrational frequencies at a given temperature as the frequency prefactor $v_{at}$, which is calculated as follows:

$$v_{at} = \frac{\int_0^{v_{max}} n(v,T) v g(v) dv}{\int_0^{v_{max}} g(v) dv} \quad (6)$$

where $n(v,T)$ is the average number of phonons corresponding to a vibrational mode of frequency $v$ at temperature $T$ with the expression as follows:

$$n(v,T) = \frac{1}{e^{hv/k_B T} - 1} \quad (7)$$

Using the total transmission probability of H/D atoms on Cu(001) $P_{tot}(T)$, and the frequency prefactor $v_{at}$, we can estimate the rate constant $k(T)$ or hoping rate as follows [37]:

$$k(T) = v_{at} \times P_{tot}(T) \quad (8)$$

The relationship between the rate constant and the mean square displacement is [11]

$$\langle d^2(t) \rangle = l^2 k(T) t \quad (9)$$

where $l$ = 2.55 Å is the distance between adjacent hollow sites. [10,11] The diffusion of H/D atoms on the Cu(001) surface (two-dimensional system) has to consider the motions in $x$- and $y$-direction, so the relationship between the diffusion coefficient and the mean square displacement is as follows:

$$\langle d^2(t) \rangle = 4 D(T) t \quad (10)$$

From equations (8)-(10), it is straightforward that

$$D(T) = \frac{l^2 k(T)}{4} = \frac{l^2 v_{at} P_{tot}(T)}{4} \quad (11)$$

The calculated rate constants $k(T)$ and diffusion coefficients $D(T)$ for H/D atoms diffusion the Cu(001) surface are shown in Fig. 4.

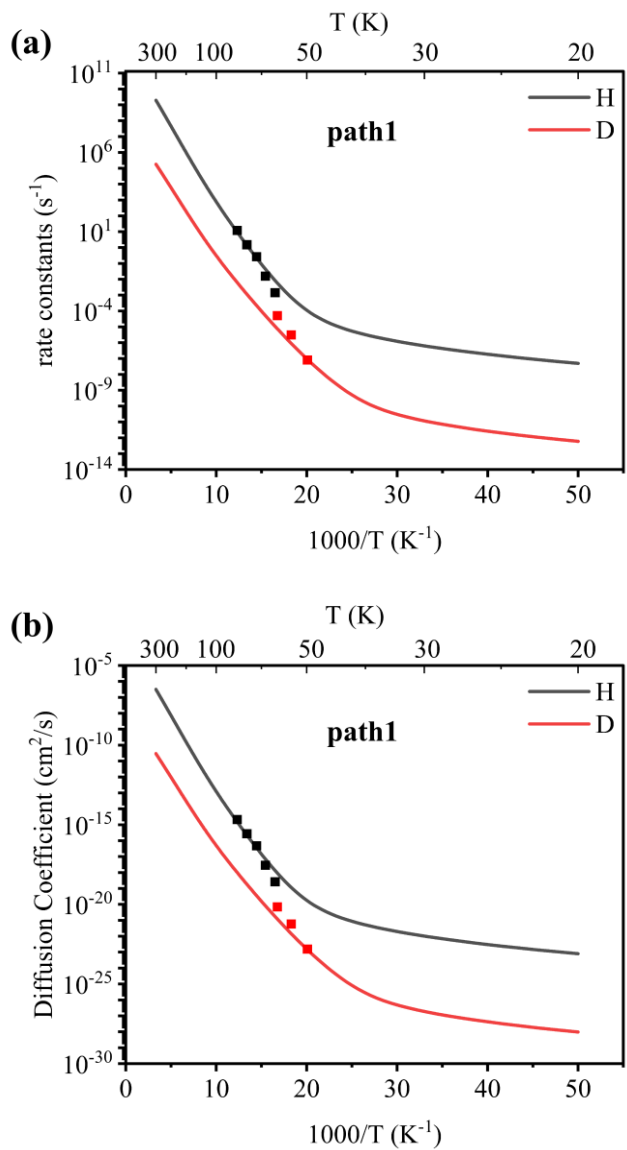

**Fig. 4** Calculated rate constants (a) and diffusion coefficients (b) of H/D on Cu(001) along path 1. The experimental data are shown as scattered squares for comparison. The black squares are experimental data for H atoms at 60 K, 65 K, 70 K, 75 K, and 80 K; the red squares are experimental data for D atoms at 50 K, 55 K, and 60 K.

Our calculations are in good agreement with the experimental results[10,11] between 50 K and 80 K. The overall good agreement justifies the validity of the temperature dependent attempting frequency evaluated by Eq. (6). In this temperature region, the classical and quantum probabilities of transmission are of the same order of magnitude and the role of quantum tunneling is less pronounced than in $T \leq 30$ K, as evidenced from Figs. 3(c)-3(d). Therefore, the diffusion of H/D follows approximately the classical Arrhenius exponential relation. Moreover, the temperature at which the transition occurs from classical to quantum-dominated behavior is approximately 55 K for H and 40 K for D, which is also in good agreement with the experimental value (~ 60 K for H, no experimental data available for D). In the case of D, our calculation predicts that the rate constant (hopping rate) is $k \sim 1 \times 10^{-11}$/s at $T = 30$ K, corresponding to a time of $10^{11}$ s for one hopping event. This is well beyond the time scale for lab measurement and helps to understand the absence of experimental data for D at $T < 50$ K. On the other hand, at lower temperatures ($T < 50$ K), the theoretically calculated rate constants $k(T)$ and diffusion coefficients $D(T)$ of H are notably smaller with comparison to the experimental measurements by more than one order of magnitude. The main reason for this discrepancy is that our calculations do not include the effects induced by the electron-hole pairs[10,11]. The experimental measurements of hydrogen atom diffusion process include the contribution of the electron-hole pair excitation, which is more influential in the low temperatures below 50 K.

## 4. Concluding Remarks

In summary, the diffusion process of H/D atoms on the Cu(001) surface was investigated based on first-principles calculations and the transfer matrix (TM) method. The minimum energy path is determined using DFT calculations and the energy path is corrected by zero-point energy which includes contribution from all phonon branches. The transmission probabilities of H/D atoms across the diffusion path are calculated by the TM method and the WKB method, with the TM method clearly demonstrating the presence of quantum interference. Finally, we calculate the rate constants $k(T)$ and diffusion coefficients $D(T)$ by introducing frequency prefactor $v_{at}$ and total transmission probability $P_{tot}(T)$. Our calculation results are in good agreement with the experimental results[10,11] at 50 K-80 K. For the subsequent work, a possible direction may be extending the above studies to the other systems and temperature regions.

**Acknowledgements**

This work is supported by the National Natural Science Foundation of China (No. 11474285, 12074382). We are grateful to the staff of Hefei Branch of Supercomputing Center of Chinese Academy of Sciences, and Hefei Advanced Computing Center for their support of supercomputing facilities. We also thank the reviewers for their reading and helpful comments on our manuscript.